\documentclass[12pt]{article}
\usepackage{amsmath}
\usepackage{graphicx,psfrag,epsf}
\usepackage{enumerate}
\usepackage{natbib}
\usepackage{url} 
\usepackage{color}
\usepackage{verbatim}
\newcommand{\blind}{0}

\usepackage[english]{babel}
\usepackage[utf8x]{inputenc}
\usepackage[T1]{fontenc}

\usepackage[colorinlistoftodos]{todonotes}

\begin{document}
\def\spacingset#1{\renewcommand{\baselinestretch}%
{#1}\small\normalsize} \spacingset{1}


\if0\blind
{
  \title{\bf Perspective from the Literature on the Role of Expert Judgment in Scientific and Statistical Research and Practice}
  \date{}
  \author{Naomi C. Brownstein\thanks{
 This work was not supported by any grant. However, the author would like to thank Jeff~Harman, Tom Louis, Tony O'Hagan, and Jane Pendergast for their helpful comments during the preparation of this manuscript.}\hspace{.2cm}
    }
  \maketitle
} \fi

\if1\blind
{
  \bigskip
  \bigskip
  \bigskip
  \begin{center}
    {\LARGE\bf Literature Review: The Role of Expert Judgment in Statistical Research and Practice}
\end{center}
  \medskip
} \fi

\bigskip
\begin{abstract}
This article, produced as a result of the Symposium on Statistical Inference,  is an introduction to the literature on the function of expertise, judgment, and choice in the practice of statistics and scientific research. In particular, expert judgment plays a critical role in conducting Frequentist hypothesis tests and Bayesian models, especially in selection of appropriate prior distributions for model parameters. The subtlety of interpreting results is also discussed. Finally, external recommendations are collected for how to more effectively encourage proper use of judgment in statistics. The paper synthesizes the literature for the purpose of creating a single reference and inciting more productive discussions on how to improve the future of statistics and science.
\end{abstract}
\noindent%

\vfill

\newpage
\spacingset{1.45} 
\section{Introduction}
As a quantitative discipline, statistics is often considered by practitioners as objective in its methods, which proliferate throughout the scientific enterprise. Basic statistical methods, including Frequentist hypothesis tests and $p$-values, are ubiquitous in academic literature. The American Statistical Association (ASA) recently warned the research community about common misuses of $p$-values in a statement \citep{asapvalstatement} that has received widespread attention in a variety of fields, even outside of statistics. 
As a follow-up, 
the ASA organized the 2017 Symposium on Statistical Inference (SSI), where statisticians and consumers of data presented research, contemplated these issues, and brainstormed possible solutions. 
Yet, 
the statistical community remains divided on explicit recommendations for researchers to improve their quantitative practice \citep{ASApvalOneYr}. 
Due to the inherently quantitative nature of the field of statistics, much of the conversation on the $p$-value statement has revolved around quantitative solutions. 
Discussions of the statistical properties of these and other remedies are included elsewhere in this special issue.

Despite perceived objectivity of the field, statisticians previously argued that uncertainty and choice abound in the scientific process, from the definitions of questions of interest to the analysis and interpretation of results \citep{gelmanhennig,goldstein2006subjective,berger1988statistical}. Hence, expert judgment is required to implement scientific research \citep{bertolaso2017evidence}. A session at SSI examined 
the 
role of judgment in statistical and scientific practice. Session participants present their joint expert opinions on the topic in \if\blind1{ANONYMOUS (2018)}\fi \if\blind0{\cite{brownstein:2018}}\fi. While preparing the paper, a wealth of related literature was collected and synthesized. The present article provides an overview of the literature on the role of expert judgment in statistics and science. 

First, section \ref{whoisexpert} defines expertise as it relates to this paper. Then, sections \ref{HTjudge},  \ref{expertBayes} and \ref{interpret} discuss the role of expert judgment in Frequentist hypothesis testing, Bayesian inference, and general interpretation. Recommendations are provided in section \ref{whatToDo} for stakeholders in education, publishing, and funding. Finally, appendices are provided for interested readers including overviews of the Frequentist and Bayesian paradigms. 

\section{What is Expert Judgment?}\label{whoisexpert}
Before discussing the role of expert judgment in science, definitions of the relevant terms are needed. \cite{Weinstein1993} provides the following definitions of expert, expertise, and expert opinion that are invoked throughout this paper: 
\begin{quote}
\begin{enumerate}
\item An individual is an expert in the `epistemic' sense if and only if he or she is
capable of offering strong justifications for a range of propositions in a
domain.\label{epistemic}
\item An individual is an expert in the `performative' sense if and only if he or
she is able to perform a skill well.\label{perform}
\item A claim is an `expert opinion' if and only if it is offered by an expert, the
expert provides a strong justification for it, and the claim is in the domain of
the expert's expertise. \label{expertopinion}
\item `Expertise' is the capacity either to offer expert opinions or to demonstrate
one or more skills in a domain, and expertise in a domain does not entail
expertise in the entire range of the domain. \label{expertise}
\end{enumerate}
\end{quote}
In brief, epistemic experts possesses deep knowledge about a field and are considered credibile by others \citep{carrier2010scientific}, while performative experts are adept at completing actions for the field. The late Stephen Hawking \citep{gribbin2016stephen} is a famous epistemic expert in physics. An olympic athlete, such as Usain Bolt \citep{gomez2013performance}, exemplifies performative expertise in their sport. The two types of expertise frequently overlap. Physicians, for instance, are both versed in fields such as anatomy and proficient at diagnosing ailments, performing medical procedures, and prescribing treatments and cures.

Expert judgment involves one or more evidence-based claims proposed by credible experts. Judgments may arise when multiple conclusions or actions are possible and supported by credible evidence. For example, multiple treatments may be possible for a condition, and two physicians may disagree on the preferred treatment plan for the same patient.
Sections \ref{statexperts} and \ref{contentexperts} relate these  definitions to statistics and science. 

\subsection{Statistical expertise}\label{statexperts}
\cite{definestats} defines statistics as ``a branch of mathematics dealing with the collection, analysis, interpretation, and presentation of masses of numerical data.'' In general, like any professional skill, statistical analysis should be conducted by practitioners who are qualified based on their knowledge of and experience with statistical procedures and principles. Example key principles include knowledge of modeling assumptions and diagnostics, as well as experience using and interpreting statistical software. 

Based on the definition of expert in \cite{Weinstein1993}, an active statistician, defined as ``one versed in or engaged in compiling statistics''  \citep{definestatistician}, is clearly considered an expert. Similar to the ability of a primary care doctor to identify and treat general health needs for their patients, statisticians are trained with knowledge and skill in choosing, creating, implementing, and validating statistical procedures for a wide variety of settings. Others can be considered statistical experts as well if they have sufficient relevant knowledge and training in statistical theory and practice.  Both types of expertise are needed for valid statistics, even if divided among multiple people. A common model is for a person with extensive training in statistical theory to supervise the study design and statistical analysis, which is executed by an analyst facile in programming.

Statistics is a broad field with numerous subfields, rendering it  impossible to know all aspects of each. Instead, analogous to  how physicians choose medical specialties, most statisticians specialize in one or more subfields, such as genetics, Bayesian modeling, or survival analysis. Classifications of subfields in statistics are found in \cite{schell2010identifying} and \cite{de2015decade}. Specialized problems require expertise in different types of analytical methods. Identifying and deciding between reasonable analytical choices requires judgment, as described throughout this paper and highlighted by others \citep{Francis2017}. 

\subsection{Content expertise}\label{contentexperts} 
While statisticians are critical members of scientific teams, interdisciplinary collaborations include experts in multiple fields. For instance, to develop a clinical trial to test an oncology drug, expertise could be required in fields such as medicine, biology, and genetics. For the purpose of this article, a content expert, or subject matter expert, has expertise in a field of science other than statistics. Content experts are key in advancing scientific endeavors, such as generating hypotheses based on knowledge of biological mechanisms, designing and supervising experiments using state of the art laboratory techniques, providing intuition on clinically meaningful values, and contextualizing statistical results.  Communication of content knowledge to the statistician is also critical, as it insight may shape the study design and analysis in subtle ways, as described in  \if0\blind{\citep{brownstein:2018}}\fi\if1\blind{ANONYMOUS (2018)}\fi.

\section{Expert Judgment in Hypothesis Testing}\label{HTjudge}
This section discusses the judgment of experts as defined in Section \ref{whoisexpert} for Frequentist hypothesis testing. 
Topics include the role of expert judgment in hypothesis testing, including finding balance between between type I and II errors, multiple comparisons adjustment, and special considerations such as orphan drugs and non-inferiority trials.

\subsection{Trade-off Between Type I and II Errors and Two Hypotheses}\label{errortradeoff}

It is well known that for a fixed sample size, there is a trade-off between minimizing false positives (type I error) and maximizing true positives (power).  Clearly, it is desirable both to have small error probabilities and high power. 
One way to minimize the probabilities of both errors simultaneously is to increase the sample size of an experiment \citep{asendorpf}. However, this solution may not be feasible, as outlined in Section \ref{stakeholders}. Instead, experts should consider their preferred balance between the two types of errors and design their experiments appropriately. 
The statistician is responsible for engaging the subject matter experts in discussions to determine the levels of each type of error that are scientifically and ethically reasonable for their study. Current methods of using judgment to balance errors in Frequentist hypothesis testing are discussed along with their implications.

\subsubsection{Minimizing Type I Error}
For most experiments, the significance level is first set. In other words, the judgment is that type I errors should be set at a certain low level, after which other considerations can be made. The most famous value is 5\%, usually chosen by convention. Rather, \cite{lakens2018justify} argues that researchers should determine their ideal level for their individual study beforehand using decision theory and report the decision and methodology in the manuscript. Unfortunately, this practice is uncommon; brief discussion is provided in section \ref{minerror}. Instead, most researchers compare to a single threshold, the customary value for which is currently under discussion. A summary of the discussion is provided.

Recent concern about the reproducibility in science focuses on the fact that interesting associations in the literature are not easily replicated in future studies \citep{ioannidis2005most}. A hypothesized cause is that the observed type I error rate may far exceed that of the nominal significance level. This hypothesis 
inspired recent recommendations to drop the most commonly used significance level further, e.g. to 0.005 \citep{benjamin2017redefine,Johnson2017,Johnson19313}.  The goal is to minimize the probability of a type I error in hopes of filtering out weak effects and increasing the reproducibility of science as a whole.

On the other hand, there is concern that the reproducibility crisis may be due to a widespread lack of power due to inadequate power calculations \citep{MARINO2017,smaldano,asendorpf}. 
Calls to lower the type I error rate have been criticized for enabling underpowered studies if no other action is taken, such as raising funding levels to accommodate increased sample sizes \citep{asendorpf,lakens2018justify}. Further, \cite{lakens2018justify} argue that the recommendation may be hurting the very cause that it champions by decreasing resources and incentives for replication studies. 

\subsubsection{Minimizing Type II Error}
The paradigm of prioritizing a fixed significance level chosen by convention only may be inappropriate for specific applications, especially if excessive harm could be associated with a {\it lack} of discovery. For example, if the question evaluates the harm of a certain environmental factor, then the precautionary principle \citep{10.3389/fpubh.2016.00107,fjelland2016laypeople} states that it is better to err on the side of caution, as failing to detect underlying harm would result in failure to reduce personal harm for individuals currently affected and future individuals. In these cases, minimizing the type II error probability is judged to be of the utmost concern. Power calculations  necessitate specification of particular realizations of the alternative hypothesis or effect size. Determination of clinically and statistically meaningful effect sizes is highly dependent on the judgment of both statisticians and content experts; guidance is provided elsewhere \citep{murphy2014statistical,ellis2010essential,lakens2013}.

The story of Love Canal, a suburban neighborhood near Niagra Falls, prioritizes minimization the harm of a type II error over that of a type I error. The general null hypothesis is the safety of the area; the alternative is an association between toxic chemicals from the Love Canal and increased cancer prevalence. The relative (bodily) harm of a type II error, declaring the area safe when it was hazardous to the residents, exceeded the (financial) harm of a type I error. 
Initially, an investigation erroneously concluded that the area was safe. The judgment of experts, namely a scientist with detailed observational data from the residents, helped diagnose the error in the initial investigation. Eventually, the area was closed. Briefly, the first study included a reasonable test to answer the wrong question. The original investigation tested the seemingly intuitive hypothesis that proximity to a chemical waste site was associated with more negative outcomes. The second investigation refined the hypothesis, namely that residents in homes near the site built over former stream-beds were more at risk than homes built over dry land. The latter test found a strong association based on sound biological principles with clear implications for evacuation prioritization. A detailed discussion of the story and the role of judgment is found in \cite{fjelland2016laypeople}.

Similarly, priorizing power, the United States Food and Drug Administration has a separate regulatory category for orphan drugs to allow new treatments to have a higher chance of adoption if no other therapeutic option is available to mitigate a rare condition \citep{braun2010emergence}. 
By definition, small populations of patients with rare conditions may make traditional recruitment targets infeasible and may require creativity in designing valid trials. Experts may design lower powered studies, increase the target significance level, consider alternative or surrogate outcomes, or even develop new methodology for small samples \citep{Parmar2016,BILLINGHAM2016e70}.  Orphan drugs trials, while usually relatively small, have sometimes, but not always \citep{orfali2012raising}, found to suffer from methodological shortcomings, such as lack of blinding or randomization \citep{kesselheim,bell2014comparison}. 
Furthermore, determination of the value of a drug is complex, based on factors such as disease prevalence, severty, mortality, morbidity, treatment benefit and safety, and cost effectiveness \citep{paulden2015value}.

\subsubsection{Minimizing Overall Error}\label{minerror}
\cite{mudge2012} recommend to consider the two types of error together and minimize the overall error rate. Two approaches include averaging the  errors themselves and calculating the overall cost based on individual error costs.
Similar approaches are applied in fields such as climate science, where the optimal $\alpha$ level is calculated via simulation \citep{kemp}. (It is worth noting that in \citeauthor{kemp}'s study, $\alpha=0.0052$, which was close to the recommendation by \citealt{benjamin2017redefine})
\cite{Mudge2017} even argue that these methods can simplify analyses with multiple hypothesis tests, which also require judgment as described in section \ref{multcomp}. More generally, \cite{grieve} analyze how overall minimization approaches relate to the likelihood principle and Bayesian decision making.
Implications of minimizing the overall error rate for research outcomes are modeled in \cite{millerulrich}.

\subsection{Expert Judgment in Alternative Frameworks: Non-Inferiority Studies}
The two hypotheses (the null and alternative) tested in standard statistical methods serve different functions, described in the appendix (section \ref{HTbasics}). \cite{sprenger2017objectivity} elaborates on the asymmetry between the two hypotheses and the inherent value judgments implicit in Frequentist methods. In non-inferiority trials, the roles of the two hypotheses are switched. (See the appendix for more details.) Non-inferiority studies require care in their design, analysis, and interpretation \citep{mauri2017challenges,fleming2011some,fleming2008current}.  
%
Examples include justification of an appropriate comparison group, 
choice of endpoint and non-inferiority margins ``based on statistical reasoning and clinical judgment'' \citep{fleming2008current}, 
and careful {\it a priori} design to handle known challenges 
\citep{mauri2017challenges,d2003non}
\cite{Rehale012594} found that statistical recommendations are unclear and those that exist aren't necessarily followed.

\subsection{Multiple Comparisons}\label{multcomp}

While testing multiple hypotheses in a single study is common, how and whether to adjust each 
test in light of the others is a non-trivial question. 
In certain settings, such as exploratory research, adjustment may be optional \citep{Wason2014,LiTalijaard,gelman2012}. Often, whether or not adjustment is needed may require judgment from a statistical expert about relationships between questions of interest (e.g. correlations or other measures of association). For example, tests comparing the effects of distinct treatments to a single control may not need adjustment, but tests comparing repeated measurements or multiple outcomes likely do \citep{Candlish2017,LiTalijaard,Wason2014}.

Even when
is adjustment for multiple comparisons is generally agreed to be important, expertise is required in specifying an appropriate procedure for a particular problem. One issue to consider first is the number of tests for which to adjust (primary questions of interest only, primary and secondary questions, all planned tests, etc.). This issue requires input from both  statisticians and subject matter experts about research priorities. The choice of adjustment method depends on the desired balance of type I and II errors, as discussed in Section \ref{errortradeoff}. Additional considerations for multiple testing are provided elsewhere \citep{LiTalijaard,Wason2014,alosh2009flexible,proschan2000practical}  

\subsection{Should a binary decision be made at all?}
The $p$-value was designed as an informal 
measure of the evidence that could cast doubt on the null hypothesis, not to inform a binary choice. 
\cite{fisher1955} objected to the decision theory framework of \cite{neyman1928use}.
\cite{christensen} describes a Fisherian view that, ``an $\alpha$ level should never be chosen; that a scientist should simply evaluate the evidence embodied in the p-value.'' The modern use of p-values arguably has lost the emphasis on scientific judgment and extrapolates far beyond their intended use. Commentaries abound on the scientific and moral implications of the abuse of the $p$-value 
\citep{Goodman1180,ziliakcult,steel2013applied,pittenger2001hypothesis,nickerson2000null,folger1989significance}. This section discusses judgment of of p-values outside of the decision-theory paradigm.

While some authors (e.g. \citealt{cumming2014}) and journals (see section \ref{stakeholders}) recommend abandoning p-values entirely, others  argue that alternatives aren't necessarily better \citep{murtaugh,macnaughton,ionides2017response,deValpine}. Instead, p-values could be used in ways other than as sole decision-makers.

First, ``borderline'' values in either direction could be judged accordingly \citep{cohenPvals} with  
replication studies encouraged to reevaluate the findings.  
Given the nominal significance level $\alpha$ is an arbitrary choice, there is little qualitative difference between $p$-values in the interval $(\alpha-\epsilon,\alpha+\epsilon)$, where $\epsilon$ is a small number much smaller than the significance level. Yet, many researchers, even statisticians \citep{mcshaneGal} reject null hypotheses corresponding to $p$-values in the lower half of the interval and fail to reject null hypotheses corresponding to $p$-values in the upper half. While intuitively, there is little difference between $p$-values of 0.049 and 0.051, papers associated with the former are far more likely to be published than papers associated with the latter. 

Additionally, p-values can be considered just one piece of evidence to be evaluated with other factors, such as study design quality and effect size \citep{spurlock2017beyond,asapvalstatement}.
In fact, the United States Supreme Court issued an opinion 
to use statistical information as evidence rather than decision tools with blunt cut-offs \citep{liptak2011supreme}.

\section{Expert Judgment in Bayesian Methods}\label{expertBayes}
To mitigate problems with Frequentist inference, a large portion of the statistical community recommends greater engagement with Bayesian methods \citep{HeldOtt,page2017beyond,goldstein2006subjective}. The Bayesian paradigm enables more intuitive interpretations that directly answer  questions of interest  \if0\blind{\citep{brownstein:2018,goldstein2006subjective}}\fi\if1\blind{\citep{goldstein2006subjective}}\fi. The vast collection of literature on Bayesian methods extends far beyond the scope of the present paper.
Appendix \ref{introBayes} provides a brief overview of the Bayesian paradigm.

It is well known that Bayesian statistics requires judgment in the choice of a prior distribution and that results may be sensitive to the choice of a prior \citep{gelmanhennig,gelman2014bayesian}. Therefore, careful consideration should be paid to the specification of the prior and its parameters. Determination of a prior requires input from both statisticians and experts in the fields of application. 
The focus of this section is an overview of prior development in Bayesian methods with emphasis on expert judgment and choice. 

\subsection{Expert Judgment in Choice of a Prior}\label{choiceprior} 

Priors can be chosen in a variety of ways.
Certain ``standard'' priors may be chosen for mathematical elegance, computational simplicity, or posterior robustness.
Classes of priors, such as non-informative priors, conjugate families, and reference priors, are described extensively elsewhere \citep{bernardo1979reference,berger1992development,syversveen,berger2009formal,deGroot,fraser2010default}. More commonly, prior distributions and parameters may be determined from pilot data or other values from the literature.  
Details on the field of empirical Bayes, involving priors derived  based on the current data, are found in chapter 5 of \cite{carlin2008bayesian}. 
\cite{little2011} describes a hybrid between Bayesian and Frequentist methods called calibrated Bayes, in which Frequentist methods help determine the prior for a Bayesian model, which is used for inference. Priors for Bayesian models can also be chosen to coincide with Frequentist procedures \citep{daita1995priors}.

\subsubsection{Expertise in Prior Elicitation}
Alternatively, 
experts 
can directly develop and calibrate a prior distribution for their project. 
The process, called elicitation, is detailed in numerous places, such as \cite{Morgan7176,o2006uncertain}, \cite{garthwaite2005statistical} and even in this special issue \citep{ohag:2018}. In brief, a facilitator carefully works with content experts to quantify their judgment about parameters of interest and either directly uses or transforms the elicited densities into prior distributions for the model. Elicitation, which falls into the subjective Bayesian paradigm 
\citep{goldstein2006subjective}, 
is used in a variety of fields, such as 
clinical trials, environmental modeling economics and ecology  \citep{mason2017development,krueger2012role,o2012probabilistic,ELE:ELE1477,martel2009eliciting}. 


\cite{johnson2010methods} urge examination of elicitation methods for validity and reliability. \cite{Morgan7176} details uses and abuses of elicitation, and \cite{heitjan2017commentary} provides recent criticisms. While bias may be induced if elicitation is done poorly \citep{kynn2008heuristics},  elicitation can be used to estimate and mitigate bias \citep{turner2009bias}. In fact,  expert judgment and elicitation is critical to modern practice in environmental modeling \citep{krueger2012role},

\subsection{Model Checking in the Bayesian Paradigm}
Regardless of the type of prior chosen for modeling, \cite{gelman2013philosophy} and \cite{sprenger2017objectivity} stress the importance of checking that the observed data matches the prior well enough for the application and considering alternative action if it diverges in key ways. For another example and discussion of implications for clinical trials when the prior doesn't match the observed data, please see \if1\blind{ANONYMOUS (2018)}\fi\if0\blind{\cite{brownstein:2018}}\fi.


\section{Expert Judgment in Interpretation}\label{interpret}

Interpretation of results involves both knowledge of the statistical modeling process and broader experience with the subject matter. For example, to understand the results of a t-test of the null hypothesis that means in two populations are equal, one must first understand the basics behind hypothesis testing in general as well as assumptions and considerations specific to the two-sample t-test, such as normally distributed populations or large samples. Misinterpretation at this stage, such as thinking a $p$-value means something it doesn't, or otherwise wrongly acting on a basic definition (confidence intervals, odds ratios, areas under ROC curves, etc.), are straightforward to correct with training in statistical literacy, as discussed in section \ref{education}. Of course, interpretation requires more than simply correctly applying definitions. Contextualizing results requires content expertise within the scientific field. Both types of expertise are required for other aspects, such as planning an analysis and evaluating potential bias, to name a few. 
This section focuses on examples including 
post-hoc difficulties with the planned analysis and missing data. Additional biases in interpretation of scientific evidence are examined in \cite{kaptchuk2003effect}.

Choices abound in designing a study, from the experimental design to the statistical analysis and interpretation; these choices require scientific judgment. (Please see \if1\blind ANONYMOUS,  2018 \fi\if0\blind\citealt{brownstein:2018} \fi for details). 
Ideally, content experts and statisticians should jointly define and prioritize scientific questions and plan statistically valid,  practical, and interpretable methods to answer them. Examples requiring judgment include decisions on the number and type of primary questions of interest, defining practically significant effect sizes on which to base sample size calculations,  and whether and when to conduct interim analyses.

A statistical analysis plan (SAP) is a formal document written prior to a study that describes the planned protocol.  \cite{Adams-Huet818} provide guidance for clinicians to work with statisticians in writing SAPs. 
Detailed SAPs are important to safeguard teams from changing their analyses after seeing the data \citep{finfer2009publish}. Such changes bias results and reduce reliability and validity. 

Even with a well-written SAP, difficulties may arise. For instance, interpretation of the SAP for adjudication of complex potential events may be difficult. In one example, a single complex case in a clinical trial required an extensive investigation to interpret the statistical analysis plan properly \citep{gibson2017academic}. In fact, the 
status of the single patient determined whether or not the $p$-value fell below the a priori defined significance level!  
When possible, SAPs can include 
strategies for dealing with missing or ambiguous data, such as sensitivity analyses, multiple imputation.

Another area which may necessitate careful statistical consideration is in the presence of excessive missing data. While a full review of missing data is outside of the scope of this paper, 
readers may consult 
\cite{little2014statistical}. Briefly, analysis with missing data requires consideration of potential relationships between missing observations and the variables in the study. Standard methods can be complicated if missing data arises in non-random ways. As an illustrative example, the gold standard diagnosis of temporomandibular disorders (TMD) requires an invasive examination by an expert dentist \citep{JOOR:JOOR2090}; missing data arises when subjects who are suspected to be new cases fail to complete the examination \citep{slade2013summary}. 
The research team showed that the proportion of missing data was large, the assumptions of standard methods were violated, and new methods were required for modeling in the presence of the missing data \if1\blind{(ANONYMOUS, 2013; ANONYMOUS, 2015)}\fi\if0\blind{\citep{SIM:SIM6604,bair2013study}}\fi. 
Recommendations, detailed further in section \ref{whatToDo}, include experimental protocols to check for and minimize missing data during data collection, and {\it a priori} plans (including existing methods or development of new methods) to handle analyses with missing data.




\section{Recommendations from the Literature}\label{whatToDo}
Commentary on the role of statistical inference in the reproducibility crisis includes calls to action for members of the scientific community, including educational reform, improved standards for publication and funding and incentives for scientists.
Section \ref{fixteaching} details recommendations for statistics education, especially regarding teaching statistical inference to students not planning to become statisticians. Section \ref{teachstatisticians} focuses on how to train statisticians with an eye toward improving reproducibility. 
Sections \ref{funding}, \ref{stakeholders}, and \ref{groups} synthesize recommendations for stakeholders to facilitate better research practice for all fields.

\subsection{Educational Interventions}\label{education} 
The need for  better statistical training in many fields has been discussed \citep{Ogino1039,sorensen2010prognosis,peng2015reproducibility}. 
Other fields are recognizing the importance of detail at each step of the research progress \citep{shipworth2018designing}. Indeed, \cite{crane2018statistics} claim that statistical training can ``[empower] scientists to make sound judgements.'' Educational programs are discussed for scientists inside and out of statistics.

\subsubsection{Educational Reform: Teaching Statistics for Practitioners}\label{fixteaching}

Despite the tendency for misuse among practitioners \citep{Colquhoun171085,gardenier} and even statisticians \citep{mcshaneGal}, basic statistical inference procedures and the accompanying $p$-values remain ubiquitous. One facilitator 
of the presence of significance tests is their central role in introductory courses  and ease of implementation in statistical software \citep{searle,dallal1990statistical}. In fact, the ``cook-book'' approach to statistical analysis commonly taught in introductory courses \citep{GIGERENZER} tempts researchers to apply basic procedures 
without investigating model diagnostics \citep{steel2013applied} or considering the appropriateness of the 
procedures for particular applications. 

\cite{crane2018statistics} and \cite{BrownKass} stress that students in statistics courses should learn how to think critically. Furthermore, ``statistical practice is complex, relying on nuanced principles honed through years of experience'' \citep{crane2018statistics}. In writing their guidelines to weed researchers,  \cite{Onofri} make the following statement, ``We would like to reinforce the idea that statistical methods are not a set of recipes whose mindless application is required by convention; each experiment or study may involve subtleties that these guidelines cannot cover.''

Paralleling the discussion for research, curriculum reform discussions are ongoing \citep{Gould2018,parkTeaching,gould2018challenge}. 
Example remedies for teaching students to use better judgment include increased focus on effect sizes and confidence intervals \citep{calin2017after,fritz2012effect}, greater attention to the Bayesian paradigm
\citep{page2017beyond}, and practical demonstrations \citep{parkTeaching,mitchell2018teaching}. 

Beyond the classroom, recommendations for biomedical scientists include programs on experimental design for both students and mentors, emphasis on methodology in journal clubs, developing continuing education materials as new methods arise, and increasing training for peer reviewers \citep{casadevall2016rigorous}. It is especially important to better train peer reviewers in all fields to enforce statistical standards in published literature \citep{peng2015reproducibility}. Scientists can pursue additional formal statistical training while in graduate school or postdoctoral programs through funded mechanisms, such as NIH T32 training grants. However, the  implementation of these specialized programs requires care to prevent trainees from simply learning a small amount of statistical tools and more confidently using them even when the tools are inappropriate for future problems \citep{gelfond2011principles}. 
\cite{PON:PON3046} recommend both improved statistical training for content experts and continued collaboration with biostatisticians throughout the research process.

\subsubsection{Improved Training for Quantitative Scientists}\label{teachstatisticians}
Statisticians serve as collaborators in a wide variety of fields, some of which strongly recommend or even require a statistician on the team \citep{obremskey2011getting,PON:PON3046}. Consequently, proper training for statisticians and other quantitative science is paramount. Recognizing the widespread need for biostatistics in research and health practice as early as the 1960s, the National Institutes of Health (NIH) created biostatistics training programs \citep{hemphill1961probable}. A more modern overview of training opportunities for biostatisticians is included in \cite{kennedy2007opportunities}

Often working in fields outside of their primary educational training, statisticians should strive to continually learn about the subject-matter problems at hand \citep{BrownKass}.
\cite{BrownKass} encourage graduate students in statistics to pursue a second program of study in joint or separate programs focused on the subject matter of their future research. Training opportunities, such as NIH K25, are available for this purpose, as described in \cite{CNCR:CNCR29033}. Formal training in a second field may not be feasible for all statisticians. However, applied practitioners could instead be encouraged to concentrate their collaborations in a small number of areas where they can slowly gain deeper familiarity with the scientific content, rather than to consult in a large number of disciplines where such depth is infeasible.  In fact, this recommendation is often communicated in informal settings to junior faculty members with foci in applied statistics.

In addition, statisticians should hone their soft skills, especially listening, communication, and leadership skills \citep{gibson2017academic,califf2016pragmatic}. 
While listening and communication are obviously essential for collaboration, leadership training can provide statisticians with the tools and confidence to continually and actively shape the statistical validity of a project from its inception. In her March 2018 president's corner article of AMSTAT news, \cite{lavange2018building} reiterated the importance of leadership in biostatistics and announced the ASA's vision for a new leadership initiative. 
As an example of the addition of soft skills to the curriculum, the biostatistics department at the University of North Carolina at Chapel Hill previously offered a course in leadership \citep{lavange2012preparing}. 
\subsubsection{Using Science to Develop Interventions for Scientists}
A final idea about educational interventions is to study the research process scientifically and use the findings to improve training. \cite{leek2017five} summarized the state of affairs with a few apt quotes:
\begin{quote}
``The root problem is that we know very little about how people analyse and process information... We need to appreciate that data analysis is not purely computational and algorithmic — it is a human behaviour... We need more observational studies and randomized trials — more epidemiology on how people collect, manipulate, analyse, communicate and consume data. We can then use this evidence to improve training programmes for researchers and the public.'' 
\end{quote}
To this end, in addition to actions described in section \ref{funding}, funders could seek studies of scientific judgment with components to develop training materials based on the findings.

\subsection{Funding Agencies}\label{funding}
Due to the fact that funding is often necessary for research and highly valued or even required for promotions, funding agencies should take an active role in defining standards. 
To this end, the NIH defined guidelines for increased rigor, transparency and reproducibility \citep{hewitt,collins2014nih}. 
The NSF followed with guidelines to encourage data sharing and citation \citep{stan2015nsf}. 
The ASA defined its own recommended actions for funding agencies to improve reproducibility, including funding mechanisms for  training in reproducible research methods, development of reproducible software, replication studies \citep{broman2017recommendations}. Actions by funding agencies should catalyze more widespread adoption of the recommendations for scientific rigor. 

\subsection{Journal Guidelines}
\label{stakeholders}
Changes in publication standards percolate to the scientific practice of authors, as opined by \cite{asher} over two decades ago. In fact, this special {\it TAS} issue stems from high profile journal articles and rules. 
Namely, after \cite{nuzzo2014scientific} bemoaned the ritual, often thoughtless, use of $p$-values, 
the editors of {\it Basic and Applied Social Psychology} banned $p$-values in  manuscripts submitted thereafter \citep{TrafimowMarks,Trafimow}. Recently, in {\it BMC Medical Research Methodology}, hypothesis testing has been ``discouraged'' \citep{hanin2017statistical}. This section considers journal guidelines and their implications.

A recent set of journal guidelines for more careful publication practices \citep{McNutt679} has been adopted by hundreds of journals thus far \citep{hewitt}. Similarly, {\it BMC Medical Research Methodology} \citep{hanin2017statistical} advises authors that ``Health care decisions [should be] based on... a combination of statistical and biomedical evidence.''  \cite{sorensen2010prognosis} propose that statistical training for journal editors could enable editors to better enforce statistical standards for publication in the journals that they oversee. 

Journal guidelines include calls for transparency, including sharing of data and code, such as with pre-submission check-lists to ensure key aspects are considered \citep{McNutt679}. Other key aspects include reporting and justification of sample size calculations, randomization and blinding procedures, and inclusion and exclusion criteria \citep{asendorpf}. Finally, emphasis can shift from inference to parameter estimation. \citep{asendorpf,McNutt679}. These guidelines encourage authors to confront, acknowledge and justify the scientific judgments used in their studies and enable others to examine or reproduce their findings. The recommendations will likely improve rigor throughout science. Yet, implications from guidelines of some journals \citep{hanin2017statistical,Trafimow} to avoid sharp thresholds for statistical inference is unclear, underlying a large portion of this special issue.

\subsubsection{Transparency}\label{transparency}
Because statistics requires choices throughout the process, statistics has elements of subjectivity and objectivity, and \cite{gelmanhennig} argue that the discussion of relative subjectivity and objectivity is distracting from solutions for best research practices, including transparency. Instead, paralleling the aim of journals to strive for transparency, \cite{gelmanhennig} detail principles for which to strive during analysis, including, among others, acknowledgment and investigation of multiple perspectives, impartiality in decision making, and transparency in reporting. In brief, they argue that statistical judgment is inevitable, and therefore, the those judgments should be detailed and shared for future evaluation. \cite{gibson} stresses that statisticians should communicate ``relative advantages'' and disadvantages of each choice.

It has been shown that not only are descriptions of methods sections insufficiently detailed, but the quality of the reporting, methodology, and analysis are not necessarily associated with increased visibility \citep{nieminen2006relationship}. In addition, poor descriptions or unwillingness to share data may be associated with author concern over the robustness of results, especially for borderline results \citep{wicherts2011willingness}. This may be related to the need discussed in section \ref{education} to increase statistical literacy broadly across other fields and professions. Statisticians can serve key roles as co-investigators by co-writing methods sections with sufficient detail for reproducibility, as peer reviewers who are specially trained to look for detail in reporting of study design, methods, and results, and as mentors and teachers who help others improve the completeness of the writing and peer review skills.

The ASA statement urges that ``Proper inference requires full reporting and transparency.'' Others similarly recommend increased transparency  of study design and analysis \citep{gelmanhennig,Greenland2016,peng2015reproducibility,McNutt679,asendorpf}, 
which would allow readers of the literature to  evaluate findings in light of the strength of the methodology. 
There are even journals dedicated to reporting raw data 
\citep{DIB,WANG201485}, but data sharing is not yet widely adopted \citep{vasilevsky}. 
Calls for transparency in research mirror calls nearly a decade ago for transparency in journalism 
\citep{weinberger}. However, others point out that too much {\it a priori} fixation and reporting of analysis plans stifles discovery \citep{poole2010vision}, and transparency alone cannot overcome poor data quality \citep{gelmanEthics}.

\subsection{Interaction Between Groups}\label{groups}
Clearly, changes to scientific and statistical standards will require coordination and buy-in from multiple groups concurrently, ensure that changes improve scientific practice as a whole \citep{ioannidis2014make,sorensen2010prognosis}. \cite{asendorpf} include an excellent summary of recommended changes for these groups. Importantly, \cite{smaldano} point out that incentives of various groups may conflict, such as funding agencies requesting publications from their grantees but not necessary checking the publication quality. 
As another example of potential misalignment of incentives, a seemingly simple solution advocated in \cite{asendorpf} to increase sample sizes across science would improve reproducibility by improving power overall and likely subsequently increase the proportion of reported discoveries that are true. On the other hand, as the cost of each additional study participant may be large, increasing sample sizes may pose financial difficulties for already strained funding bodies. If total budgets were not increased, then requiring studies to be larger (and thus more expensive) could result in fewer projects funded. Such a consequence would negatively impact individual researchers who would face increased competition for comparatively fewer grants. Additional discussion of guidelines and their implications for various groups is found in \cite{WILLIAMS2018197}.

\section{Conclusion}
There is broad consensus that rigorous statistical practice and teaching require a mixture of technical knowledge, communication skills, and the ability to interpret data. 
The present paper highlights the less frequently discussed presence of choice and judgment in scientific and statistical practice. The universe of potential methods for analysis is large, and deciding among the options is challenging, even for experts \citep{gelman2014we}. Properly synthesizing expert judgment along with quantitative tools is critical for high quality evidence-based research. 
Failure to appropriately capitalize on statistical and scientific judgment could further erode public trust in science \citep{Speigelhalter,SALTELLI20175} and policy \citep{sutherland2015use,weinberg2012science}.  More importantly, scientists have an ethical obligation to conduct valid statistical analyses, which eventually have broader societal impacts \citep{shmueli2017research,zook2017ten,gelfond2011principles}. 

Rather than argue for a single solution to the problem of improving quantitative practice, this paper collects in a single document many of the thoughts on expert judgment and recommendations in statistics. (In a related piece submitted to this special issue, a strong statement on expert judgement in science with example applications by participants in the SSI is provided in \if\blind1{ANONYMOUS (2018)}\fi \if\blind0{\citealt{brownstein:2018}}\fi.)  
The present article gives the reader a starting point to understand the vast literature related to expert judgment in statistics. Viewpoints  are highly diverse, indicating that there is much more work to be done toward developing concise, unified recommendations for improved methods. 
The author intends for the present paper to facilitate ongoing discussions of expert judgment and recommendations to improve statistical practice in the twenty-first century and beyond. 

\bibliographystyle{agsm}
\bibliography{mybib}

\section{Appendix: The Basics and Philosophy of Hypothesis Testing}\label{HTbasics}
Hypothesis testing compares two hypotheses, called the null and the alternative. Two paradigms are presented for comparing the hypotheses using data collected in scientific experiments. Other overviews are found elsewhere, e.g. \cite{greenlandPoole}.

\subsection{The Frequentist Paradigm}\label{FreqBasics}
In Frequentist hypothesis testing, the null hypothesis is considered the default, assumed to be true unless the data casts doubt on it. The alternative hypothesis is often what a content expert may hope to conclude is true by casting doubt on the null hypothesis. For example, the developer of a new cancer treatment may compare changes in tumor size for patients randomized to either their new treatment or a placebo. In this case, the null hypothesis is that average change in tumor size is identical for patients randomized to both interventions, and the alternative hypothesis is that the change in tumor size differs based on whether the patents were randomized to the new drug or a placebo. Put another way, the drug developer hopes to convince regulatory bodies that their drug is effective by casting doubt on the default theory that the effect of the drug is identical to the effect of a placebo.

If one assumes that the two (mutually exclusive) hypotheses exhaust reasonable possibilities, then exactly one of the hypotheses should be true and the other should be false. The practitioner either rejects or fails to reject the null hypothesis based on the evidence in the experiment. While the hope is that the conclusion aligns with the (unknown) truth, two errors can take place. The first, called a type I error, occurs when the null hypothesis is true, but the practitioner rejects the null hypothesis. The second, called a type I error, occurs when the alternative hypothesis is true, but the practitioner fails to reject the null hypothesis. Type I errors can be conceptualized as false positives, while type II errors can be considered false negatives. The significance level is the probability of a type I error. Power, equal to the probability of the complement of a type II error, is the probability of (correctly) rejecting the null hypothesis when the alternative hypothesis is true.

For a fixed sample size, type I and type II errors are inversely related (when one increases, the other decreases), and the appropriate balance should be decided carefully.
Typically, type I error is set at an appropriate value to minimize the occurrence of false discoveries, which are typically considered worse than missing a true discovery. The sample size for the experiment is then set based on both the significance level (e.g. 5\%) and a desired level of power (e.g. 80\%) under a specified realization of the alternative hypothesis.  Alternatively, often in the presence of extremely limited resources, power is calculated  based on the fixed significance level and the maximum sample size considered feasible. 

The significance level is arbitrary, but 5\% has been the most frequently adopted value by tradition \citep{ziliakcult}. 
The idea is that the probability of falsely rejecting the null hypothesis, e.g. concluding that an association of interest in the study is present in the population when in reality no such association exists, needs to be minimized. Otherwise, not only will the original findings mislead the research community, but future research will be designed based on the previous (incorrect) findings.

As detailed in \cite{motulsky2014intuitive}, one can conceive of the Frequentist hypothesis framework as similar to a jury evaluating the evidence in a criminal trial: the defendant is presumed by default not to be guilty of the crime and is only deemed to be guilty of the crime if there is sufficient evidence of guilt beyond a reasonable doubt.  In the court room, wrongful convictions are considered worse than failing to convict a person who committed a crime in the presence of insufficient evidence. Similarly, type I errors are typically set at lower rates (e.g. 5\%) than type II errors (10\% or 20\%). Additional connections between statistics and law and examples for teaching are provided in \cite{byun}.

It is important to note that the null and alternative hypotheses serve different functions and are not interchangeable. A common misconception is to interpret a failure to reject the null hypothesis as ``accepting the null hypothesis.'' Critically, $p$-values do not quantify the evidence in favor of the null hypothesis, analogous to the fact that criminal trials never claim to prove the innocence of a client and merely deliver verdicts without convictions as as ``not guilty'' \citep{motulsky2014intuitive}. In some cases, the order of the hypotheses may need to be flipped, such as for non-inferority trials, where the goal is to disprove the notion that a generic compound differs from the original version in favor of the alternative that the two compounds are sufficiently similar. Thus the framework, analysis, and interpretation of non-inferiority studies differ markedly from the standard hypothesis testing framework.

The body of the paper discusses an inverse relationship between type I and type II errors when the sample size is fixed. Exacerbating the trade-off is the fact that most research aims to answer more than one question of interest, and therefore, a significant portion of the literature includes more than one hypothesis test. The family-wise error rate, defined as the probability of having at least one type I error among a fixed number of tests, is known to be higher than the nominal significance level of any single test, as shown in \cite{LiTalijaard}. 
In fields, such as genomics, with a large number of planned hypothesis tests, adjusted methods aim to control the false discovery rate, defined as the proportion of the null hypotheses that are true among all tests for which the null hypotheses is rejected \citep{benjamini1995controlling}. It is important to note that family-wise error rates and false discovery rates are not equivalent. Calculation of the false discovery rate utilizes Bayes' rule and necessitates a judgment on the proportion of null hypotheses that are true.

\subsection{The Bayesian Paradigm}\label{introBayes}
As described in Section~\ref{FreqBasics}, the Frequentist paradigm answers questions about the data assuming that one of the competing hypotheses is true. In particular, a $p$-value quantifies  the evidence an experiment produces {\it assuming that the null hypothesis is true}. In reality, even after the experiment is complete, the research team cannot definitively answer the question of whether or not the null hypothesis is true. It is intuitively more desirable to quantify and compare the posterior probabilities of each hypothesis conditional on the data from the experiment. The Bayesian paradigm provides methodology for this framework.

Consider the example from Section~\ref{FreqBasics} of a new oncology drug. Under the Bayesian paradigm, a quantity of interest is the posterior probability of the null hypothesis that the drug and placebo change tumor size by amounts that are nearly identical (up to a range) based on the data from the patients in the trial and the previously hypothesized efficacy of the drug. Similarly of interest is the posterior probability that the drug reduces the tumor size by a greater margin than the placebo conditional on the observed data. The drug developer hopes to convince regulatory bodies that their drug is effective by directly calculating the posterior probability, based on the results of the trial, that there is a meaningful underlying difference in treatment response between their drug and a placebo.

Another key difference between Bayesian and Frequentist methodology is dependence of Bayesian posterior probabilities on a prior distribution. The prior specifies the probability that the null hypothesis is true (or false). Another quantity of interest in this paradigm is the Bayes factor, which is the ratio of the likelihood of the data under one hypothesis to the likelihood of the data under the other hypothesis. The philosophy behind the Bayes factor is detailed in \cite{morey2016philosophy}.

More generally, Bayesian methods involve specification for the parameters of interest. In the presence of nuisance parameters, i.e. parameters that are not of interest themselves but must be defined to answer the question of interest, this endeavor may be especially challenging \citep{fraser2010default,tibshirani1989noninformative}. Further details on the Bayesian paradigm and how to implement Bayesian models may be found in a variety of sources, such as textbooks by \cite{gelman2014bayesian} and \cite{carlin2008bayesian}.

\end{document}